\documentclass[journal ]{new-aiaa}
\usepackage[utf8]{inputenc}
\usepackage{textcomp}

\usepackage{graphicx}
\usepackage{amsmath}
\usepackage[version=4]{mhchem}
\usepackage{siunitx}
\usepackage{longtable,tabularx}
\usepackage{authblk}

\title{Portable Laser-Pumped Rb Atomic Clock with Digital Circuits}

\author[1]{Qiang Hao\thanks{wlsichuang@163.com}}
\author[1]{Shaojie Yang}
\author[1]{Peter Yun}
\author[1]{Jun Ruan}
\author[1]{Shougang Zhang}

\affil[1]{Key Laboratory of Time Reference and Applications, 
National Time Service Center, Chinese Academy of Sciences, 
Xi’an 710600, China}

\begin{document}

\maketitle
\begin{abstract}
Reducing the size and complexity of high-performance timekeeping devices is an ever-growing need for various applications, such as 6G wireless technology, positioning, navigation and timing (PNT), Internet of Things (IoT), and ultrafast spectroscopy. This work presents a distributed feedback (DFB) laser-pumped Rb atomic clock, which features extraordinary frequency stability, small size and low power consumption. The DFB laser head employs a built-in isolator with a linewidth of approximately 1 MHz. For complete optical pumping of the atoms in the absorption cell, the laser beam is expanded to a diameter of 10 mm by using an optical diffuser-based beam expander. The physics package is based on a magnetron microwave cavity and surrounded by two layers of magnetic shielding. The overall volume of the optical system combined with the physics package is 250 cm$^3$. The proposed atomic clock is also designed to operate at a low temperature, whose absorption cell is maintained at 323 K. Benefiting from the lower Rb atom density, the excited atoms present a long population relaxation time of 5.8 ms. The frequency synthesizer and frequency-locked loop are implemented by digital circuits. The short-term stability of the atomic clock is measured to be $1.8\times10^{-12}\tau ^{-1/2}$ (1-100s). Our achievement paves the way for practical application of the laser-pumped Rb atomic clocks. 
\end{abstract}

\section{INTRODUCTION}
Time Synchronization and timestamp are essential for numerous emerging applications such as 6G wireless technology, navigation and timing (PNT), Internet of Things (IoT), and ultrafast spectroscopy. These applications increasingly demand timekeeping devices that are more precise and lower in size, weight, and power (SWaP) \cite{2021GPS, Kitching, Optical2024, Mercury}. The atomic clock is a stable and precise timekeeping device based on the intrinsic quantum transitions of atoms, ions or molecules. To achieve a portable atomic clock, large number of studies focus on the coherent population trapping (CPT) technology by using the vertical cavity surface emitting laser (VCSEL). It has been well known as the chip-scale atomic clock (CSAC) \cite{NIST,2005vanier}, which features ultra-low power consumption and small size. The typical frequency stability of the commercial available CSAC is $10^{-10}\tau ^{-1/2}$ level \cite{Japan}. Meanwhile, several miniaturized Rb atomic frequency standards (RAFSs) have also been developed for a higher precise timekeeping capability, which present typical frequency stability of $10^{-11}\tau ^{-1/2}$ level \cite{2021GPS}. 

\begin{figure}[!t]
\centering
\includegraphics[width=3.5in]{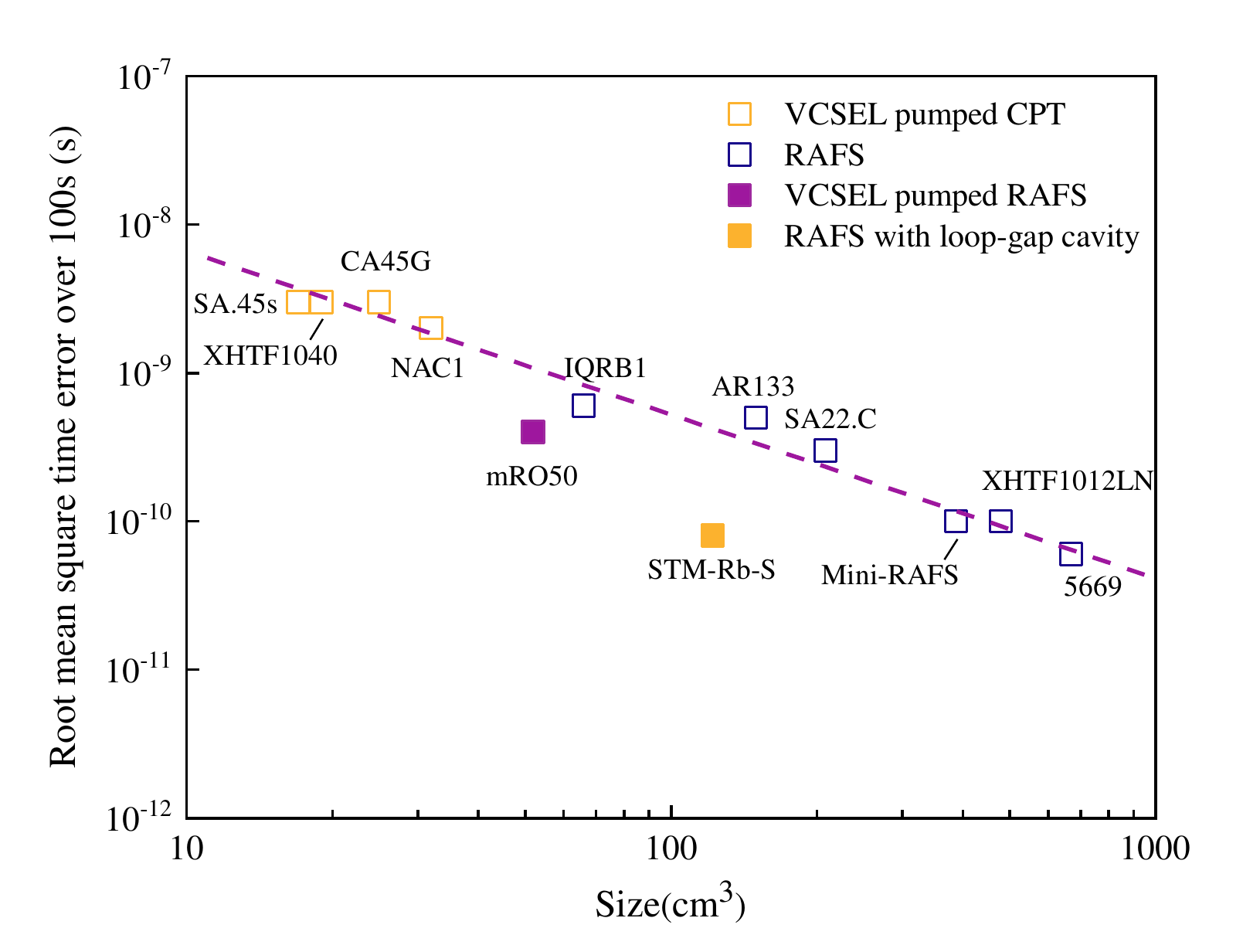}
\caption{Time prediction error of some commercial portable atomic clocks. The solid line represents a linear fit, showing an averaging performance dependence on the size. SA.45s: Microchip SA.45s, XHTF1040: Chengdu Spaceon XHTF1040, CA45G: Femtolocking CA45G, NAC1: Accubeat NAC1, mRO50: Safran mRO50, IQRB1: IQD IQRB1, AR133: Accubeat AR133, STM-Rb-S: Chengdu sync STM-Rb-S, SA22.C: Microchip SA22.C, Safran Mini-RAFS,   XHTF1012LN: Chengdu Spaceon XHTF1012LN,  5669: FEI 5669.}
\label{fig1}
\end{figure}

Assuming that the initial time and frequency offsets are zero, in the case of the white frequency modulation noise, the optimum time error of prediction is $\tau \sigma_y (\tau )$, where $\sigma_y (\tau )$ is the Allan variance of the atomic clock, $\tau$ is the sample interval \cite{Allan1978}. Considering most of the portable atomic clocks are deployed at the user terminal and their long-term stability could be disciplined by global navigation satellite systems(GNSSs) or other higher-performance ones \cite{Discipline}, we take $\tau=100$ s. Figure 1 shows the time prediction error against the size of some typical commercial atomic clocks, which are smaller than 1000 cm$^{3}$.

It is observed that the clock performance presents a strong dependence on the size \cite{Review}, which is because reducing the size of atomic clocks generally introduces more white frequency noise from electronic components and decreases the signal-to-noise ratio of the physics package. It is worth noting that there are two exceptions, which are the VCSEL-pumped RAFS (Safran mR050) \cite{Safran2023} and the RAFS equipped with a loop-gap microwave cavity (Chengdu sync STM-Rb-S) \cite{sync}.

The commercial VCSEL-pumped RAFS was reported with  $4\times10^{-11}\tau ^{-1/2}$ stability and 51-cm$^3$ volume. This exception is because of the improved optical pumping efficiency and reduced shot noise from VCSEL \cite{Liu}. The improvement contributed by the loop-gap cavity is due to the highly oriented microwave field, which allows more atoms to be involved in the optical-microwave double resonance \cite{mileti2012, Mei2024}. Although VCSEL has shown advantages over the discharge lamp, it has notable limitation on the linewidth (50 MHz). In contrast, the RAFS that is pumped by a laser of 1 MHz level linewidth (hereinafter referred to as “laser-pumped RAFS”) has demonstrated significantly improved frequency stability of $1.3\times10^{-13}\tau ^{-1/2}$ \cite{Bandi2014}. However, there is a significant challenge to miniaturize such a kind of atomic clock \cite{M1987, Vanier2007}.

In this work, we introduce a portable laser-pumped RAFS with high frequency stability, compact size and digital circuits. The overall volume of our laser system and physics package is 250 cm$^3$. The operation temperature of our physics package is around 323 K, showing potential for low power consumption application. The short-term stability is measured to be $1.8\times10^{-12}\tau ^{-1/2}$(1-100s).

In addition, we carry out a quantitative comparison of the shot noise limit of laser-pumped and lamp-pumped RAFSs. The shot noise limit could be expressed as $\sigma_{shot}(\tau ) \propto \frac{\Delta \nu}{ (S/N)} \tau^{-1/2}$, $\Delta \nu$ and $S/N$ are the linewidth and signal-to-noise ratio of the double-resonance signal. High coherence of the laser enables a higher $S/N$, however, it also leads to enhanced saturated broadening, which is difficult to theoretically predict because of its spatial dependence \cite{INRIM2009}. 
\section{Atomic clock design}
The configuration of a laser-pumped RAFS is similar to that of a lamp-pumped one, but the isotopic filter cell and the discharge lamp are replaced by a laser system. A laser-pumped RAFS primarily consists of three parts: physics package, laser system and electronic controller. 

\subsection{Laser system}

Low-noise laser diodes mainly include extended cavity diode lasers (ECDLs), distributed feedback (DFB) lasers, and distributed Bragg reflector (DBR) lasers. The ECDL has a linewidth of about 100 kHz, but its mode hopping and sensitivity to acoustic noise remain significant concerns for practical application. On the contrary, DFB and DBR lasers have broad mode-hopping free tuning range and exhibit a linewidth of around 1 MHz. Due to the buffer-gas collision and Doppler broadening, the absorption linewidth of the absorption cell is about 1 GHz. Therefore, 1 MHz linewidth is adequate to a short-term stability of $10^{-13}\tau ^{-1/2}$ level \cite{Bandi2014}. The  DBR laser could emit higher power than the DFB laser. In contrast, the DFB laser tends to have a lower threshold current \cite{OL2021}. In our case, a few mW power is needed and the threshold current should be as low as possible to reduce the power consumption. Therefore, the DFB laser is a better option. 

\begin{figure}[t]
\centering
\includegraphics[width=3.5in]{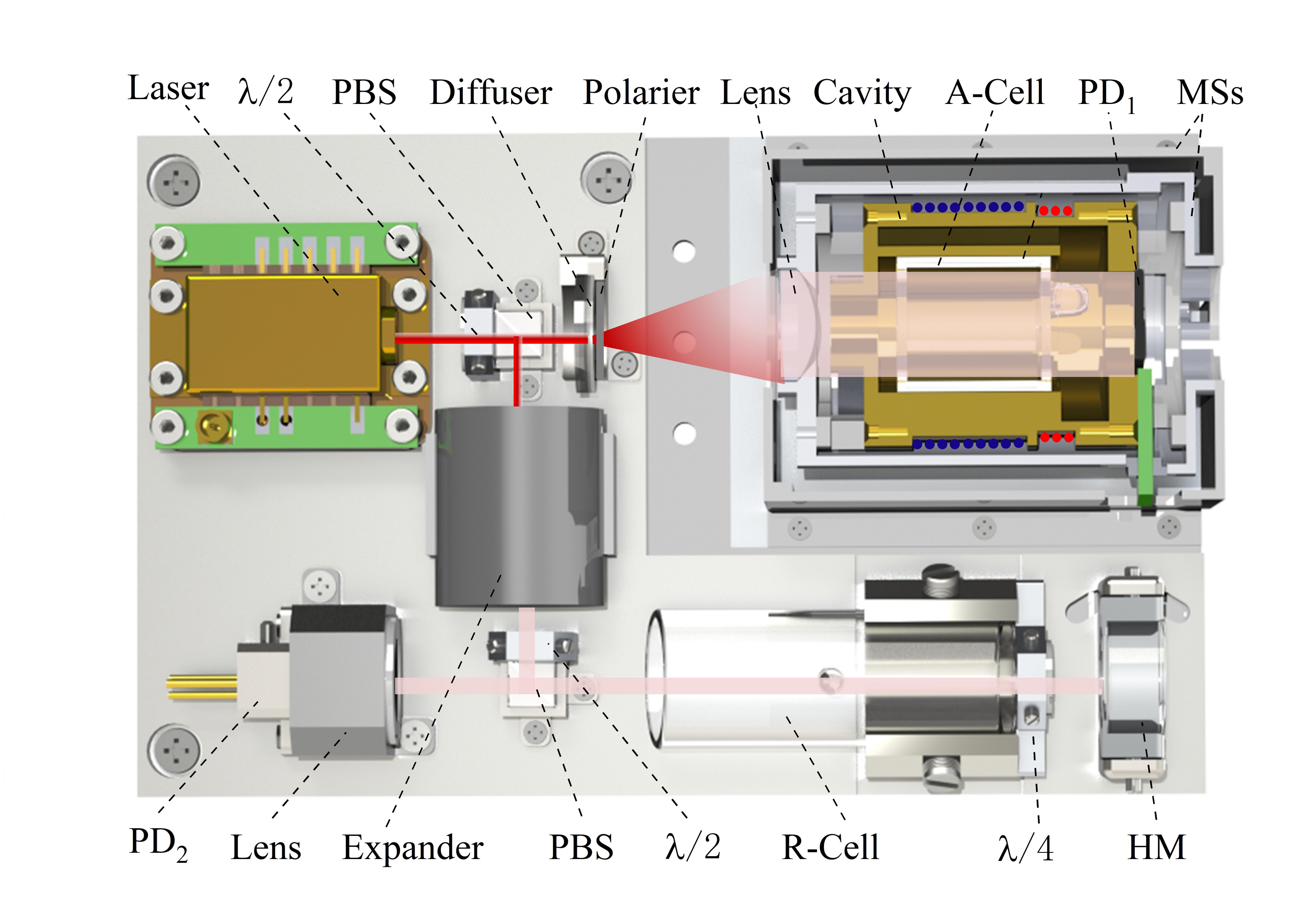}
\caption{Drawing of optics system and physics package. The width and length are 72 mm and 115 mm, respectively. The height of the physics package is 35 mm and the height of the optics system is 25 mm, the total volume is about 250 cm$^3$. A-cell: absorption cell, PD: photodetector, MSs: magnetic shields, PBS: polarizing beam splitter, HM: half-reflecting mirror. }
\label{fig2}
\end{figure}
 
 \begin{figure}[ht]
\centering
\includegraphics[width=2.5in]{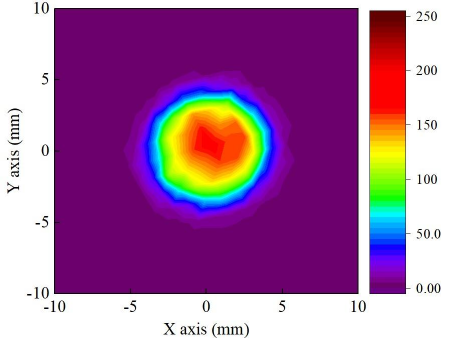}
\caption{ Power distribution of the beam after diffuser-based expander, which is measured at the position of the photodetector (PD$_1$).}
\label{fig3}
\end{figure}  
  As an integrated semiconductor laser, DFB laser is insensitive to acoustic noise and humidity, but a beam collimation is required due to its edge emission. Meanwhile, an optical isolator is necessary to block the unwanted feedback into the laser diode. Here, we use a  DFB laser that integrates the laser chip,  thermoelectric cooler, thermistor, monitoring photodetector (PD), collimator lens and micro isolator in the hermetic butterfly package (EYP-DFB-0780-00040-1500-BFW11-0005). The threshold current of the DFB laser is 32 mA.\par
\begin{figure}[ht]
\centering
\includegraphics[width=2.5in]{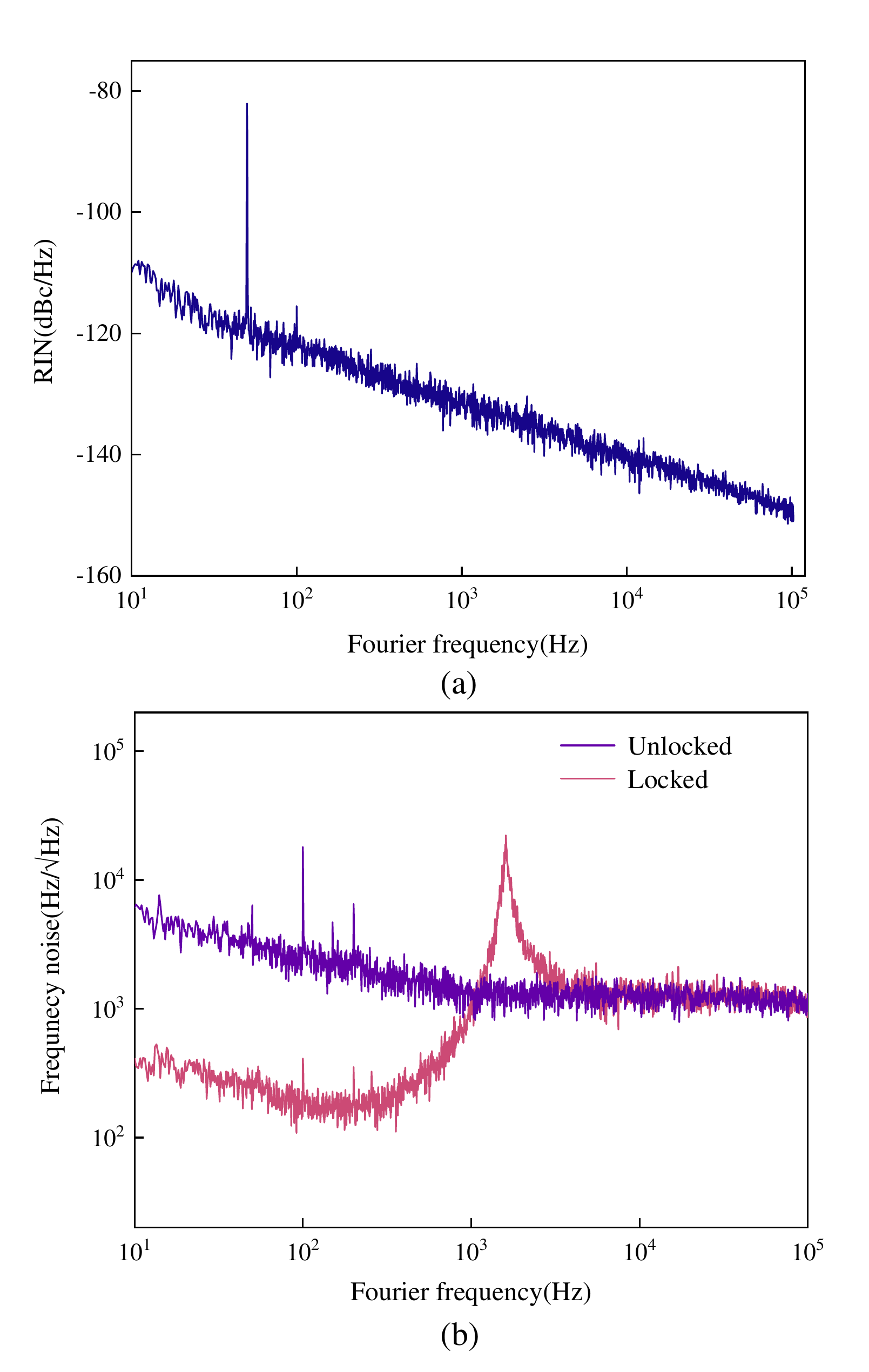}
\caption{Relative intensity noise (a) and frequency noise (b) of the pumping light. The relative intensity noise and the locked frequency noise are measured when the laser frequency is locked to Doppler-free signal $|5S_{1/2}, F_g=2 \rangle \rightarrow|5P_{3/2}, F_e=1,3\rangle$}.

\label{fig4}
\end{figure}
 The output beam needs to be enlarged by 10 times in diameter from 1 mm to match the absorption-cell diameter. The widely used expanders are based on the lens group which include Keplerian and Galilean beam expanders. The Galilean expander consists of a negative and a positive lens, enabling a short overall length. Even in an extreme case, when using a plano-concave lens of -6 mm focus length (FL) and a plano-concave lens of 60 mm FL, the length of the beam expander is larger than 50 mm. In addition, the expanded beam pointing is sensitive to the position and tilt of the lenses due to the short focus length and small dimension. As shown in Fig. 2, we design a highly integrated beam expander by using an optical diffuser and a lens. The diffuser has a diameter and thickness of 12.7 mm and 1.1 mm. The transmission and the diffuse angle of scattering are 80\% and 15$^\circ$, respectively. A lens of 20 mm FL is placed after the diffuser and integrated into the physics package. 

The design of the optical system is as follows: the light emitted from the laser diode is split into two branches by a polarizing beam splitter (PBS). The main part with 3 mW power passes through the diffuser and a polarizer, which is used to control the intensity and purify the polarization of the pumping light. The light is then collimated by the lens and illuminates the atoms in the absorption cell. The other branch is used for saturated absorption spectroscopy (SAS). The laser beam is expanded to 3 mm in diameter by a Galilean expander and then directed to the reference cell, which is 12 mm in diameter and 40 mm in length. The probe light of the SAS is focused with an 8 mm FL lens and detected by a PD in 1.2 mm diameter, which has a response speed of 100 MHz. All the optical elements were selected with a small footprint. For example, the PBSs are 5 mm cubes and the wave plates are  6 mm in diameter.

Fig. 3 shows the light beam image after the beam expander. The result presents a Guass distribution and the horizontal and the vertical beam diameters (1/$e^2$) are 9.1 mm and 8.9 mm, respectively. So the beam circularization of the expanded light is 0.98. Considering the distribution of the polarized atoms in the central absorption cell is larger than that of the edge, the Gaussian beam could contribute to a higher signal-to-noise ratio than the uniform distribution \cite{INRIM2009}.

Another important consideration is the power consumption. The temperature and current of the DFB laser are set to 305 K and 52 mA so that it can emit 5 mW light at the wavelength of 780.24 nm. As is well known, the Rb discharge lamp generally operates at 383 K, so the laser-pumped approach could potentially reduce power consumption. The laser current driver is based on the Hall–Libbrecht method \cite{laser}, the modulation and demodulation signal is generated by a digital signal generator. The modulation frequency and peak-to-peak amplitude are 2 MHz and 6 mV, respectively. The reference cell works at room temperature without temperature control.

\begin{figure}[ht]
\centering
\includegraphics[width=3.5in]{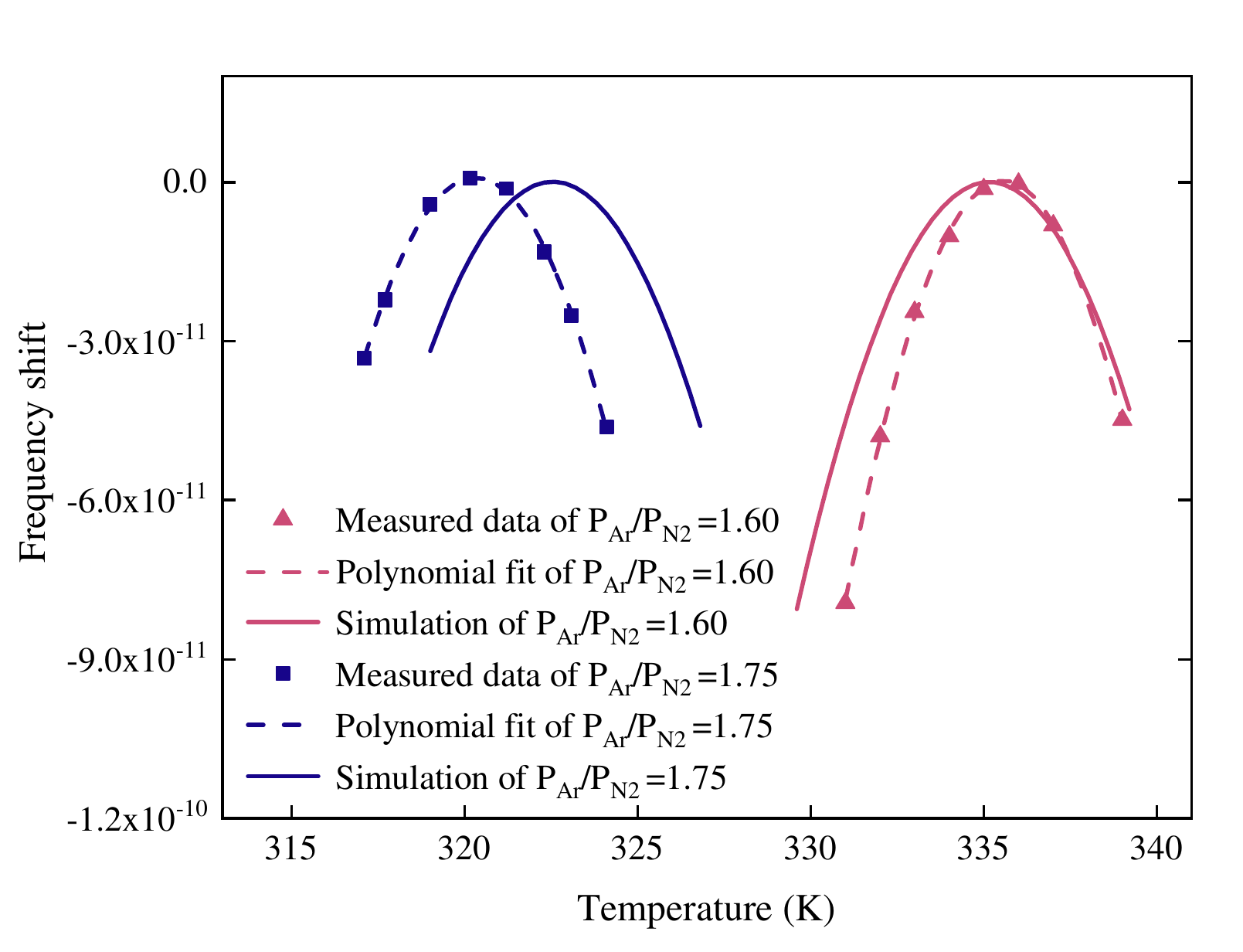}
\caption{Experimental and calculated temperature-induced frequency shift, where the constant collision shift is removed.}
\label{fig4}
\end{figure}
We measure the laser intensity noise with a fast Fourier transform spectrum analyzer \cite{2017yun}. Since the laser system is designed without active optical components, the intensity noise of the pumping light is identical to that of the laser diode. The relative intensity noise (RIN) is found to be -120 dBc/Hz at the atomic clock modulation frequency of 70 Hz. The impact of RIN on short-term stability is evaluated to be $3.1\times10^{-13}\tau ^{-1/2}$. Additionally, the laser beam is 3-fold magnified to improve the signal-to-noise ratio of SAS, the slop of the error signal is 0.65 $\mu$A/Hz of Doppler-free signal $|5S_{1/2}, F_g=2 \rangle \rightarrow|5P_{3/2}, F_e=1,3\rangle$, and the frequency noise is 195 Hz/Hz$^{1/2}$ at the offset frequency 70 Hz.

\subsection{Physics package}

The physics package is designed to produce a stable frequency-discrimination signal. It requires a highly orientated microwave field, long-lifetime excited Rb atoms, large attenuation of the external magnetic field, a constant temperature, and a homogeneous static magnetic field. As is shown in Fig. 2, the microwave cavity is a monolithic magnetron cavity that holds an absorption cell with an internal diameter $d_0$=13 mm and an internal length $l$=14 mm. The resonant mode of the cavity has been confirmed to be TE$_{011}$-like through the Zeenman transition spectrum. The buffer gases in the vapor cell are the Ar and N$_{2}$ mixture of 25 torr for ﬂuorescence quenching, Dicke narrowing \cite{Dicke} and low-temperature sensitivity.  Solenoids provide a static magnetic field of 10 $\mu$T. The physics package has two layers of magnetic shielding, which are expected to have an axial magnetic field attenuation greater than 30 dB \cite{Nie}. A 10 mm square PD (PD$_1$ in Fig. 2) is used to detect the transmitted light. The overall physics package has a volume of 62 cm$^3$, which is similar to the one previously developed in our group except for an additional layer of magnetic shield and an embedded PD \cite{2024}.

The pressure and mixture ratio of the buffer gases play a critical role in the temperature sensitivity of the physics package. The buffer-gases collision shift could be expressed as \cite{Vanier1982}
\begin{equation}
\label{equ1}
\nu (T)=\nu_ 0+{P}({\beta^{'}+\delta ^{'}}(T-T_0) + {\gamma ^{'}}{(T-T_0)^2})
\end{equation}
where $P$ is the total gas pressure, $\beta^{'}$is constant collision shift, $\delta ^{'}$ and $\gamma ^{'}$ are the weighted linear and quadratic temperature coefficient of the buffer-gas mixture, respectively. $T_0$ is the reference temperature. Pressure ratio P$_{Ar}$/P$_{N2}$ =1.6 has been widely used in laser-pumped RAFS \cite{Almat2020, Qiang2024}. The point of zero-temperature coefficient is measured to be 335 K, which agrees well with the calculation, as plotted in Fig. 5.
\begin{figure}[!t]
\centering
\includegraphics[width=3.5in]{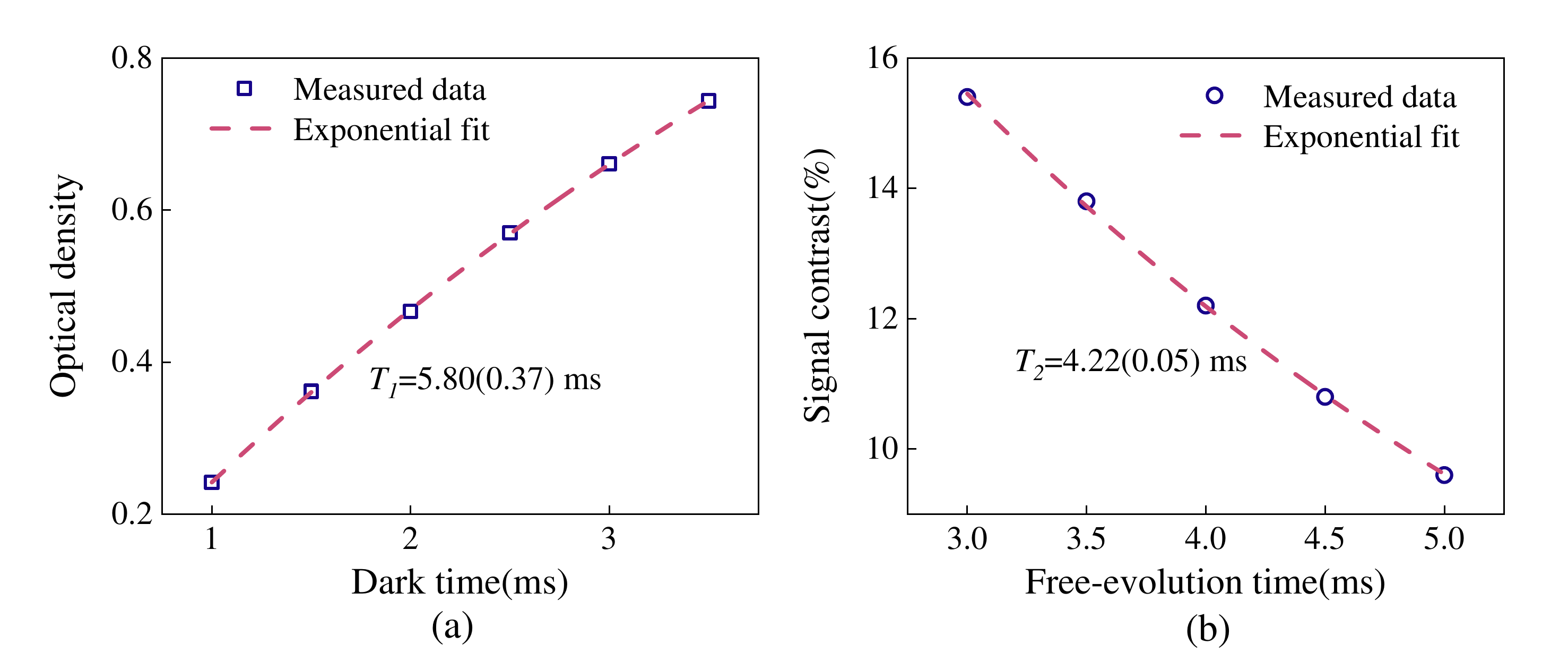}
\caption{The population (a) and coherent (b) relaxation time under temperature of 323 K, where optical density is defined by -$ln(I_t/I_0)$, $I_0$ and $I_t$ are the transmitted and incident light intensity, respectively.}
\label{fig6}
\end{figure}

 Due to heating the physics package is one of the main contributors to the power consumption of the portable Rb atomic clock, the absorption cell is designed to be operated at a low temperature. According to (1), the theoretical value of zero-temperature coefficient will reduce to 323 K when the mixture ratio is P$_{Ar}$/P$_{N_2}$=1.75. As is shown in Fig. 5, measured data confirms that the point of zero-temperature coefficient is 320 K. A frequency shift of 3914 Hz is observed, which agrees well with the predicted result of 4033 Hz, suggesting precise filling of the buffer gases. We therefore interpret the 3 K difference between the experiment and calculation as the measured error of the second-order coefficient. In this sense, the point of zero-temperature coefficient for the ratio of 1.75 shows a greater calculated error than that of the ratio of 1.6, whose point of zero-temperature coefficient is much closer to T$_0$ (333 K). Since the buffer gas ratio could be easily adjusted as 1.72 to achieve a zero-temperature coefficient at 323 K. We still maintain the absorption cell at 323 K in this study for high signal-to-noise ratio consideration.

The uncertainty principle indicates the measurement uncertainty of energy levels depends on atoms' relaxation time. By using the pulsed optically pumping technique, the population and coherence relaxation are evaluated by the Franzen method \cite{Franzen} and the decay rate of the Ramsey signal \cite{INRIM2021}. As is plotted in Fig. 6, the population and coherent relaxation are exponentially fitted to be 5.80 ms and 4.22 ms. Benefiting from the low temperature of the absorption cell, our results are much higher than those of previous studies [28], [30]. This improvement is attributed to the reduction of Rb atom density by a factor of 2.3, which greatly decreases spin-exchange relaxation. When the temperature is increased by 10 K for the same vapor cell, the population and coherence relaxation are decreased by 84\% and 42\%, respectively \cite{2024}.

\begin{figure}[!t]

\centering
\includegraphics[width=3.5in]{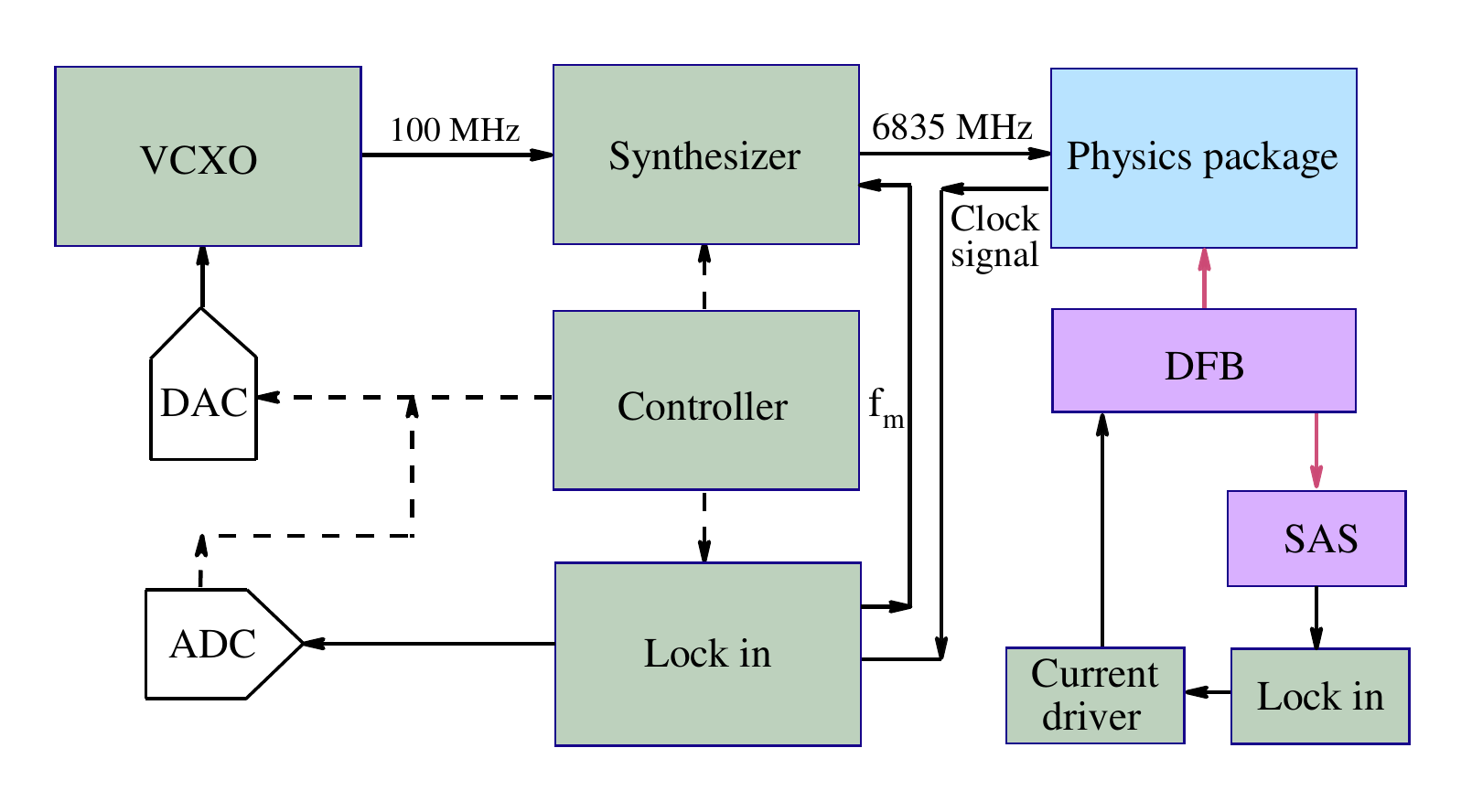}
\caption{Schematic diagram of the laser-pumped atomic clock based on digital circuits, ADC: analog-to-digital converter, DAC: digital-to-analog converter, SAS: saturated absorption spectroscopy, VCXO: voltage-controlled crystal oscillator.}
\label{fig7}
\end{figure}
\subsection{Digital circuits}
The digital controller is more suitable for portable atomic clocks than analog circuits due to its ease of integration, flexible control, and convenient diagnosis. A schematic of the atomic clock is shown in Fig. 7, in which the optical and physics package is integrated as shown in Fig. 1. The electronic controller is demonstrated by using a set of commercial digital devices. In the frequency synthesizer, a phase-locked dielectric resonator oscillator (PLDRO) generates a 6.8 GHz signal, which is referred to a 100 MHz voltage-controlled crystal oscillator (VCXO). Another branch of 100 MHz signal is down-conversion to 10 MHz by a direct digital synthesis (DDS) providing the reference to a signal generator (Rohde Schwarz SMA100A). The signal generator outputs a 34.686 MHz signal, which is frequency modulated by a modulation frequency $f_m$=70 Hz and a modulation depth of 220 Hz. The 6.835 MHz signal is produced by the 6.8 GHz and 34.686 MHz signals using a frequency mixer (Mini-Circuits ZX05-83+). The absolute phase noise of the microwave signal is measured as -106 dBc/Hz at the offset frequency of 140 Hz, which leads to an inter-modulation effect contribution of $5.0\times10^{-14}\tau ^{-1/2}$ \cite{Dick}.

A lock-in amplifier based on the digital signal processor (SR 830) is used to demodulate the transmitted light. The modulated signal and demodulated reference are both provided by the lock-in amplifier. The error signal is collected by a 16-bit analog-to-digital converter (DAC) in a multi-function data acquisition board (NI 6363). The data is then averaged and processed by a virtual proportional-integral (PI) controller. Finally, the VCXO is controlled by a 16-bit digital-to-analog converter (DAC). The sampling rate of both channels are 100 kS/s. The digital control loop also allows auto-locking and re-locking techniques without adding any hardware.
\section{Short-term stability}
The physics limit of the atomic clock $\sigma _{y}(\tau )\propto \Delta \nu/C$, where the signal contrast $C=i_0/I_t$, $i_0$ is amplitude of the double-resonance signal, $I_t$ is the background light. $C$ increases with the power of the microwave signal. However, excessive power causes saturation broadening, as is observed in Fig. 8. $\Delta \nu/C$ is observed to be the minimum when microwave power is approximately -65 dB. The frequency-discrimination curve and DR signal are plotted in Fig. 9 (a). The linewidth and signal contrast are 630 Hz and 16.5\%, respectively. The slope of the frequency-discrimination curve in the modulation region is measured to be 0.39 nA/Hz. The shot noise limit of the system can be evaluated by \cite{Bandi2014}
\begin{equation}
\label{Eq2}
\sigma _{shot}(\tau )=\frac{\sqrt{2eI_{dc}}}{\sqrt{2}  D\nu_0 }\tau ^{-1/2}
\end{equation}
where $e$ is the electron charge, $I_{dc}$ is the transmitted photocurrent at the half intensity of the double-resonance signal. Therefore, the shot noise limit $\sigma _{shot}(\tau )$ is $1.6\times10^{-13}\tau ^{-1/2}$, indicating the high potential of the proposed portable laser-pumped RAFS.

\begin{table}[htb]
\centering
\caption{Characteristics of various Rb atomic clocks }
\setlength{\tabcolsep}{1.2mm}{
\begin{tabular} {lcccccr}

\hline\hline
\textrm{Type}&
\textrm{$d_0$(mm)}&
\textrm{$l$(mm)}&
\textrm{D(nA/Hz)}&
\textrm{$I_{dc}$($\mu$A)}&
\textrm{ $\sigma _{shot}$(1 s) }&References\\
\hline
Lamp & 12 & 14  & 0.37 & 25 & $7.9\times10^{-13}$& \cite{Qiang2015}\\
 Laser& 13 & 14 & 0.39   & 1.15 & $1.6\times10^{-13}$ &This work\\
 Lamp & 18 & 18 & 1.4 & 118 & $4.5\times10^{-13}$ &\cite{Feng}\\
 Laser& 18 & 18 & 0.88 & 1.9 & $9.2\times10^{-14}$& This work \\
\hline\hline
\end{tabular}}
\label{tab1}
\end{table}

\begin{figure}[ht]
\centering
\includegraphics[width=3.5in]{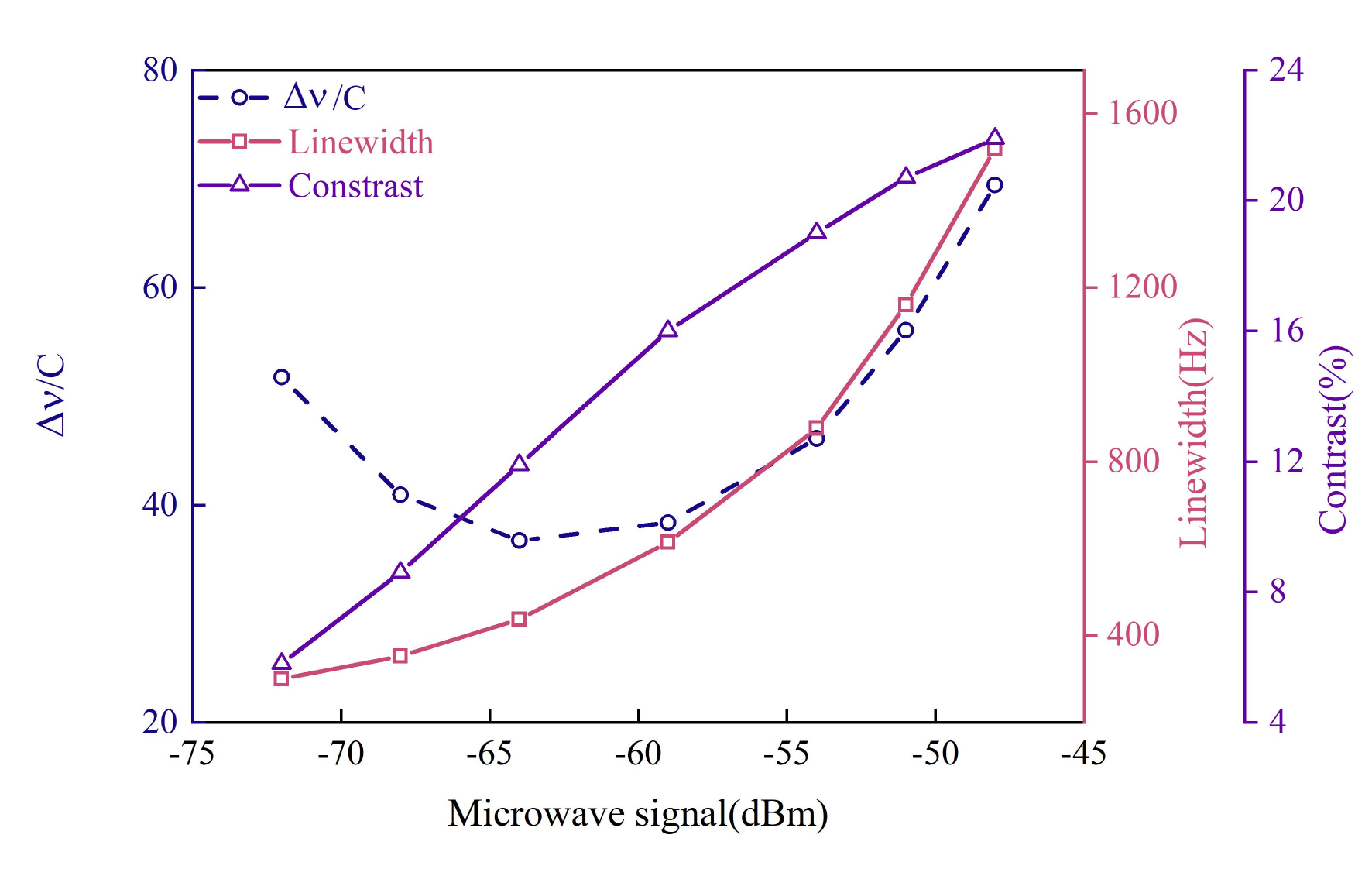}
\caption{Contrast (purple), linewidth (red) and the ratio of linewidth to signal contrast (dashed navy blue) versus microwave power.}
\label{fig8}
\end{figure}
\begin{figure}[!t]
\centering
\includegraphics[width=3in]{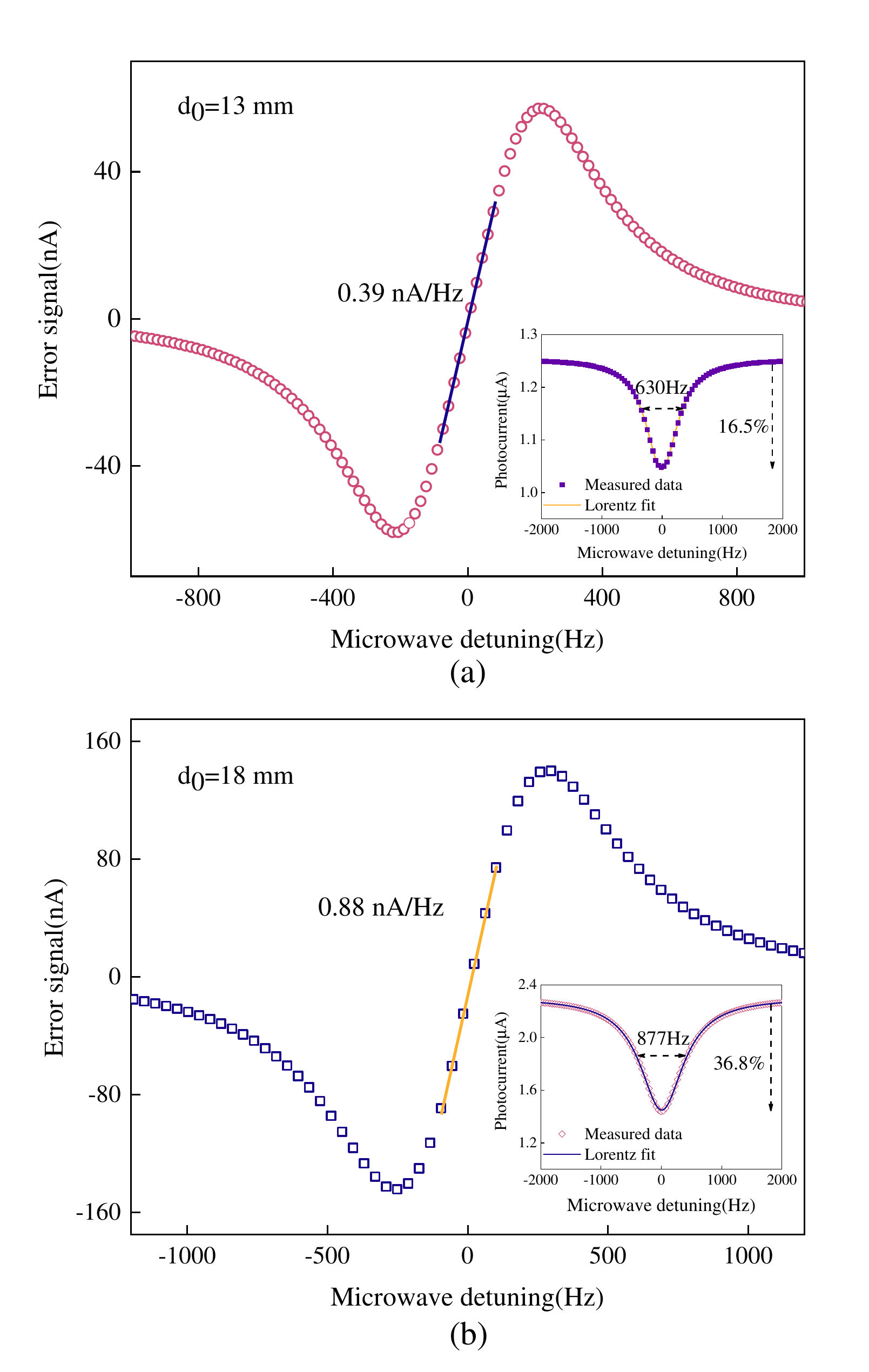}
\caption{Frequency discrimination curves of two different laser-pumped RAFSs, in which the internal diameter of the vapor cell $d_0$=13 mm (a) and $d_0$=18 mm (b). Insets are the measured double-resonance signals.}
\label{fig9}
\end{figure}

\begin{figure}[ht]
\centering
\includegraphics[width=3.5in]{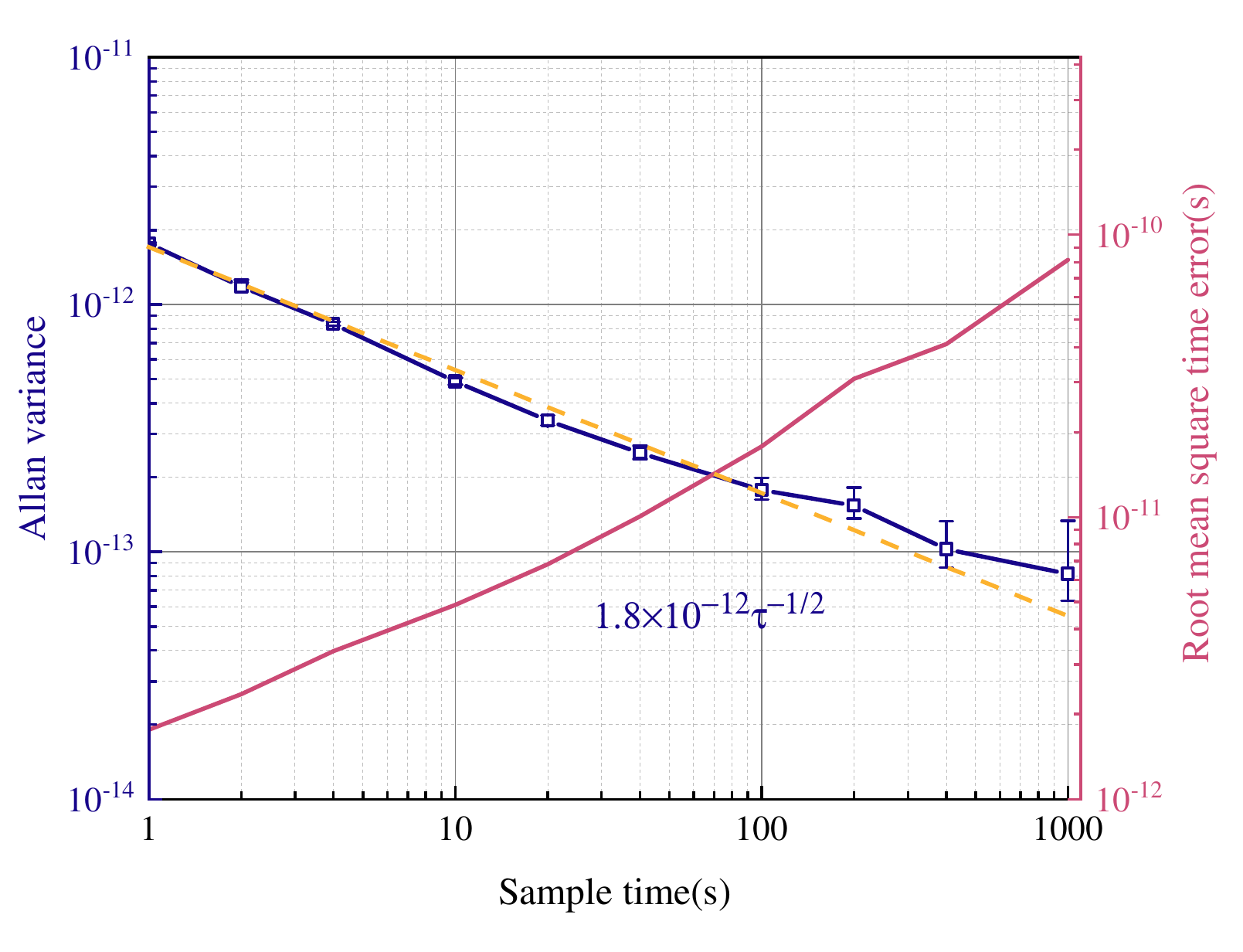}
\caption{Allan variance (navy blue line ) and the time error prediction (red line)}
\label{fig10}
\end{figure}

Quantitatively comparing the performance of laser-pumped and lamp-pumped RAFS is challenging, which requires constructing high-performance versions of both types. However, comparing our results with the previous lamp-pumped RAFS using an absorption cell with $d_0 = 12$ mm \cite{Qiang2015} is easy to be conducted. It is found that the shot noise limit of the laser-pumped approach shows 4.94-fold improvement.

Considering $\sigma _{shot}(\tau )$ is parameter-specific, we provide another complementary comparison with a vapor cell of $d_0 = 18$ mm \cite{Qiang2020}. After parameter optimization, we measure the frequency discrimination curve and double-resonance signal of the laser-pumped RAFS, shown in  Fig. 9(b). The frequency-discrimination slope and the contrast are 0.88 nA/Hz and 36.8\%, resulting in a shot noise limit of $9.2\times10^{-14}\tau ^{-1/2}$. It demonstrates 4.89-fold improvement over the lamp-pumped one using a similar vapor cell \cite{Feng}. 

According to the above two sets of comparisons, we infer that the laser-pumped approach could improve the shot noise limit by half an order of magnitude. It should be noted that all the four atomic clocks are based on loop-gap microwave cavities: the laser-pumped clocks use magnetron cavities and the lamp-pumped clocks use slotted tube cavities, which ensures that the influence of shot noise limit primarily originates from the light source. In terms of lamp-pumped RAFS, the Rb spectral lamps are filled with Xe as the starter buffer gas,  and work at 383 K. The double optical filtering technique, i.e., isotopic filtering and bandpass filter are used to increase the signal-to-noise ratio \cite{Qiang2016}. Details of the four atomic clocks are listed in Table I.

The frequency fluctuation is measured against an active Hydrogen maser, which presents one-order higher frequency stability than our atomic clock. As shown in Fig. 10, the short-term stability is measured to be $1.8\times10^{-12}\tau ^{-1/2}$(1-100s), which is mainly limited by the resolution (153 $\mu$V) of the DAC using to servo the VCXO \cite{Dong}. The time errors for the duration of 1 s and 100 s are 1.8 ps and 18 ps, respectively.

\section{Light shift}
Light shift notably limits the long-term stability, which is interpreted as the virtual transitions induced by pumping light. Compared with the discharge lamp, the DFB laser has a 1000 times narrower linewidth, resulting in the light shift being both intensity and frequency dependent \cite{M1990,Mileti2005}. The light shift is generally caused by the hyperfine light shift and the tensor light shift. In particular, for the $5^{2}P_{3/2}$ energy level of $^{87}$Rb atoms, the hyperfine splitting is unresolved due to the Doppler and buffer-gas-collision broadening, so the tensor light shift dissipates. The light shift response function can be simply expressed in term of hyperfine light shift \cite{Happer1968, INRIM2016}:

\begin{equation}
\label{Eq1}
S(\nu)\approx S_{hfs}(\nu)=\frac{\lambda^2r_0f}{2\pi hc} (\frac{Mc^2}{2RT})^{1/2}Re[\zeta ^0(aa)-\zeta^0(bb)]
\end{equation}
where $a=I+1/2$, $b=I-1/2$, $h$ is the Planck constant, $\lambda$ is the optical wavelength, $r_0$ is the electron radius, $f$ is the oscillator strength of the transition. $M$ is the molecular weight of the atoms, $c$ is the speed of light, $R$ is the gas constant, $T$ is the absolute temperature, $\nu$ is light frequency, 
\begin{equation}
\label{Eq1}
\zeta^0 (F_gF_g)=2\sum_{F_e}(2F_e+1)W^2(J_eI1F_g;F_e\frac{1}{2})Z(F_eF_g)
\end{equation}
$Z(F_eF_g)$ is the dispersion function, describing the Doppler-broadened resonance lines.

Fig. 11 plots the light-shift response function, in which the relative frequency 0 MHz refers to the optical frequency of transition of $|5S_{1/2},F_g=2 \rangle \rightarrow|5P_{3/2},F_e=3\rangle$. We can see that the light shift could be the smallest when the laser frequency is locked to the crossover resonance $|5S_{1/2},F_g=2 \rangle \rightarrow|5P_{3/2},F_e=1,3\rangle$. Fig. 11 also plots the absorption spectrum in the absorption cell and the Doppler-free spectrum in the reference cell. The buffer-gas frequency shift of the absorption cell is -190 MHz. Also in experiment, we observe the light shift when the laser frequency is locked to different Doppler-free signals, shown in Fig. 12. The intensity light shift for the Doppler-free resonance $|5S_{1/2},F_g=2 \rangle \rightarrow|5P_{3/2},F_e=3\rangle$, $|5S_{1/2},F_g=2 \rangle \rightarrow|5P_{3/2},F_e=2,3\rangle$ and  $|5S_{1/2},F_g=2 \rangle \rightarrow|5P_{3/2},F_e=1,3\rangle$ are measured to be $3.7\times10^{-11}/\%$, $1.2\times10^{-11}/\%$ and $4.5\times10^{-13}$/\%, respectively. The experimental results could be well explained the the calculation plotted in Fig. 11.  From Fig. 12, the laser frequency light-shift coefficient of laser frequency is deduced to be $2.3\times10^{-11}/$MHz, indicating that the frequency light shift could also be a large contributor for the long-term stability \cite{Qiang2024}.  

\begin{figure}[ht]
\centering
\includegraphics[width=3.5in]{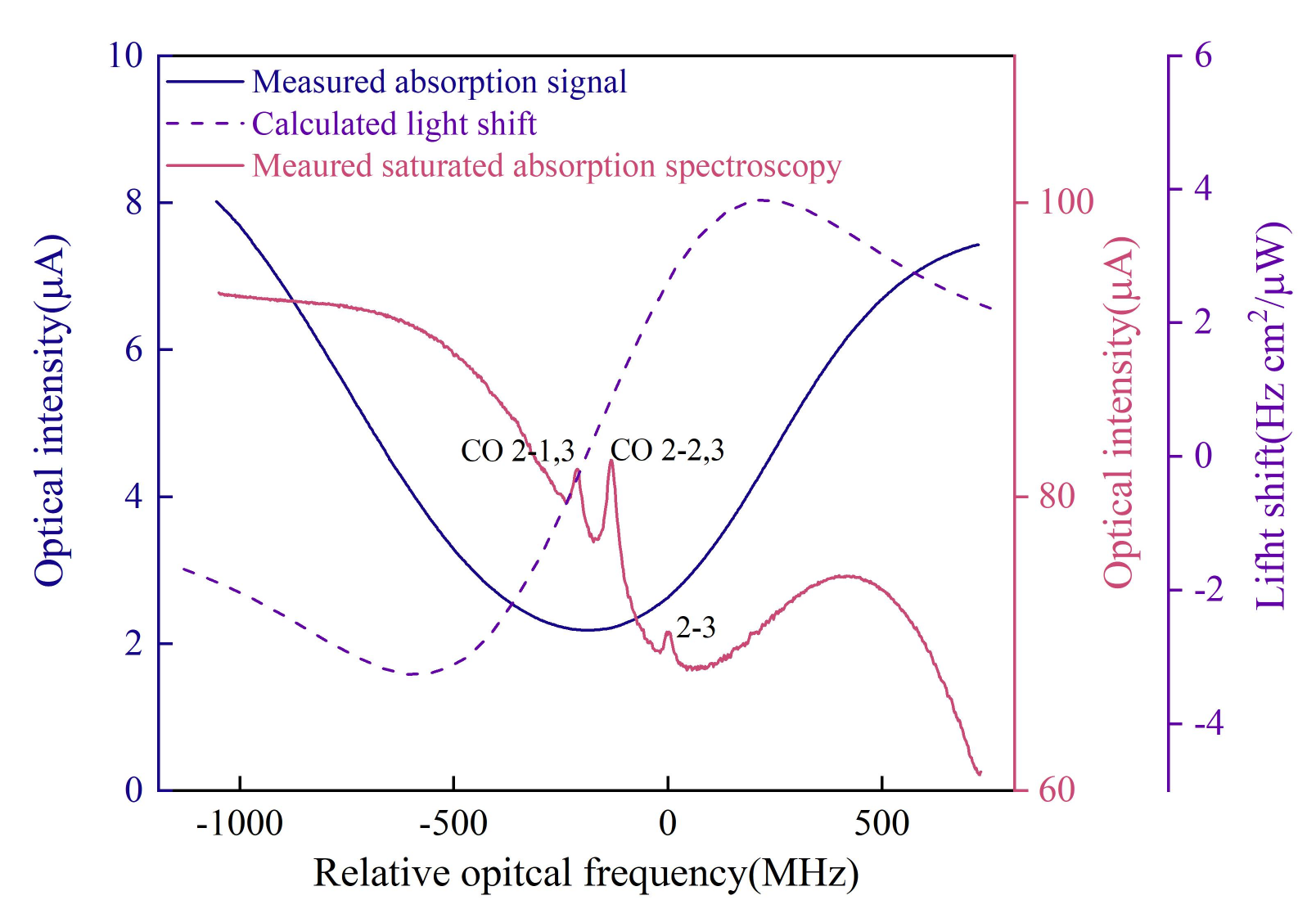}
\caption{ Measured absorption spectrum (navy blue line) and calculated light-shift response curve (purple dashed line) of the Rb atoms in the absorption cell, and saturated absorption spectroscopy observed with the reference cell. CO 2-1,3, CO 2-2,3 and 2-3 are the crossover resonances $|5S_{1/2},F_g=2 \rangle \rightarrow|5P_{3/2},F_e=1,3\rangle$, $|5S_{1/2},F_g=2 \rangle \rightarrow|5P_{3/2},F_e=2,3\rangle$ and resonance $|5S_{1/2},F_g=2 \rangle \rightarrow|5P_{3/2},F_e=3\rangle$, respectively. The relative frequency 0 MHz refers to the transition of $|5S_{1/2},F_g=2 \rangle \rightarrow|5P_{3/2},F_e=3\rangle$. }
\label{fig11}
\end{figure}

\begin{figure}[ht]
\centering
\includegraphics[width=3.5in]{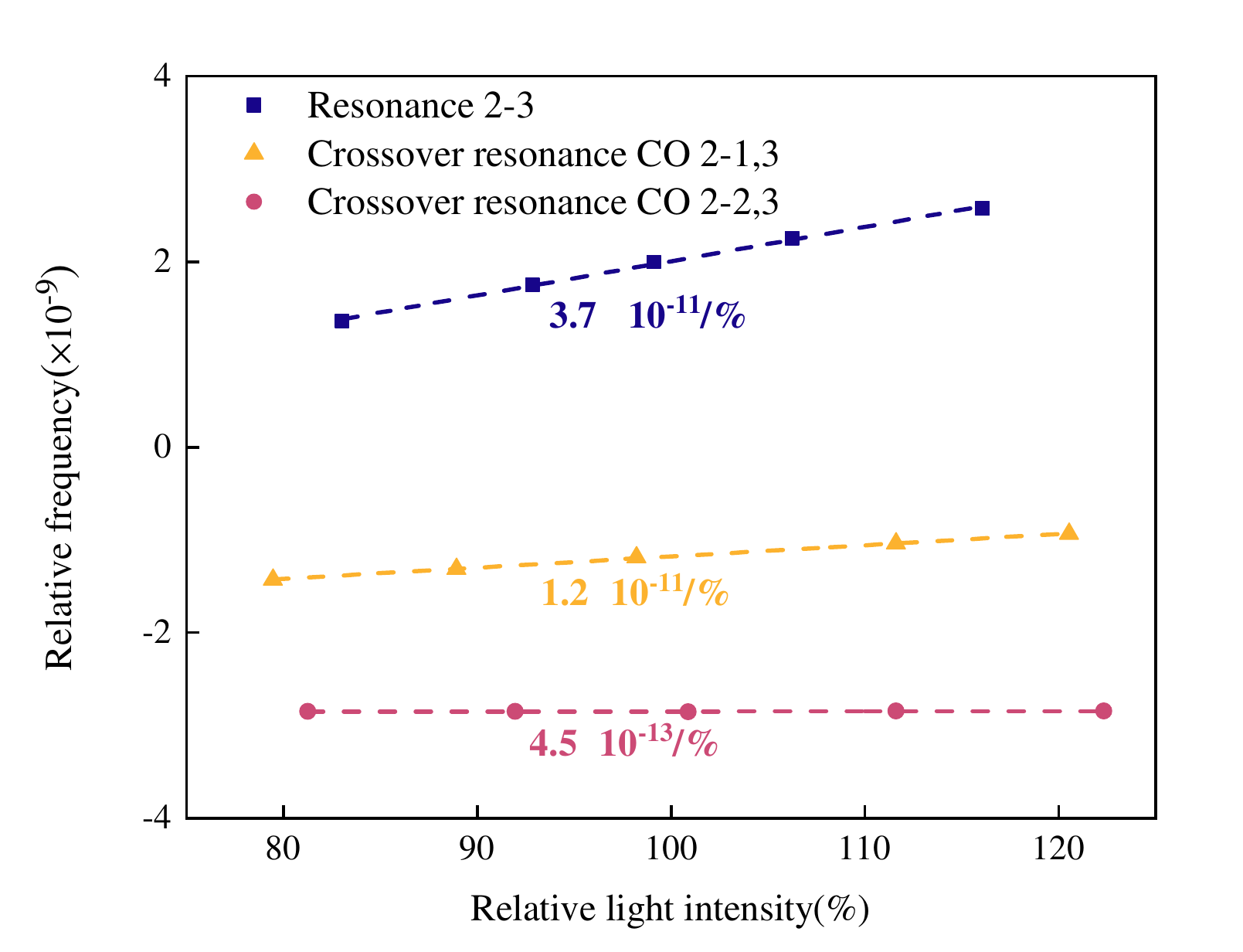}
\caption{Intensity light shift when the laser frequency is locked to various Doppler-free peaks of $^{87}$Rb D$_2$ line, where the symbols are measured data, and the dashed lines are linear fits.} 
\label{fig12}
\end{figure}
\section {Conclusions and outlook}
We demonstrate an integrated optical system combined with the physics package in volume of 250 cm$^3$, which is more simplified and compact than the previous works \cite{Mileti2006}. The laser-pumped atomic clock presents a high signal-to-noise ratio, permitting Rb vapor cell to operate at a lower temperature of 323 K. Due to the reduced spin-exchange effect, the population and coherence relaxation times are measured to be 5.80 ms and 4.22 ms, respectively. These values represent increases of 84\% and 42\%, respectively, compared to the case of 333 K. The short-term stability of the proposed atomic clock is $1.8\times10^{-12}\tau ^{-1/2}$ (1-100s).

The loop controller is implemented by using digital circuits, offering flexibility for status monitoring, parameter optimization, and drift correction. This strategy also reduces the size and power consumption, as digital signal processing requires fewer electronic components \cite{chengdian2024}. Commonly used 16-bit ADCs and DACs are employed in our demonstration, which can be integrated similarly as their usage in CSACs, charge-coupled devices (CCDs), fiber optic gyroscopes, and data acquisition boards. The frequency-locked loop of the DFB laser shared a similar principle to that of the atomic clock but with a higher modulation frequency of 2 MHz, which could be implemented by digital circuits and digital signal processing as demonstrated. The optimum microwave power is found to be -65 dB. For the future synthesizer, a phase-locked loop (PLL) could be used to build a low-power consumption and small-size microwave generator. Considering the PLL will bring in higher phase noise, a notch filter with the center frequency of 2$f_m$ could be used to reduce the intermodulation effect \cite{Deng}. Therefore, we believe that the total size of the proposed atomic clock, including integrated electronic circuits, could be less than 500 cm$^3$.

\section*{Acknowledgments}
This work was supported in part by the National Natural Science
Foundation of China under Grant 12173044

\end{document}